\documentclass[a4paper]{article}

\usepackage[margin=1in]{geometry}

\usepackage[T1]{fontenc}
\newcommand{\changefont}[3]{
\fontfamily{#1} \fontseries{#2} \fontshape{#3} \selectfont}

\changefont{ptm}{m}{n}

\usepackage{setspace} 

\usepackage{graphicx}    

\usepackage{amsfonts}
\usepackage{amssymb}
\usepackage{graphics}
\usepackage{mathrsfs}
\usepackage{color}

\newtheorem{theorem}{Theorem}[section]

\newtheorem{corollary}{Corollary}[section]
\newtheorem{lemma}{Lemma}[section]

\newtheorem{definition}{Definition}[section]

\long\def\symbolfootnote[#1]#2{\begingroup%
\def\thefootnote{\fnsymbol{footnote}}\footnote[#1]{#2}\endgroup} 

\begin{document}

%

\begin{center}
\Large \textbf{Poincar\'{e} chaos and unpredictable functions}
\end{center}

\begin{center}
\normalsize \textbf{Marat Akhmet$^{a,}\symbolfootnote[1]{Corresponding Author Tel.: +90 312 210 5355,  Fax: +90 312 210 2972, E-mail: marat@metu.edu.tr}$, Mehmet Onur Fen$^b$} \\
\vspace{0.2cm}
\textit{\textbf{\footnotesize$^a$Department of Mathematics, Middle East Technical University, 06800 Ankara, Turkey}} \\
\textit{\textbf{\footnotesize$^b$Basic Sciences Unit, TED University, 06420 Ankara, Turkey}}
\vspace{0.1cm}
\end{center}

\vspace{0.3cm}

\begin{center}
\textbf{Abstract}
\end{center}

\noindent\ignorespaces
The results of this study are continuation of the research  of Poincar\'{e} chaos initiated in papers (Akhmet M, Fen MO. Commun Nonlinear Sci Numer Simulat 2016;40:1--5; Akhmet M, Fen MO. Turk J Math, doi:10.3906/mat-1603-51, accepted). We focus on the construction of an unpredictable function, continuous on the real axis. As auxiliary results, unpredictable orbits for the symbolic dynamics and the logistic map are obtained. By shaping the unpredictable function as well as Poisson function we have performed the first step in the development of the theory of unpredictable solutions for differential and discrete equations. The results are preliminary ones for deep analysis of chaos existence in differential and hybrid systems. Illustrative examples concerning unpredictable solutions of differential equations are provided.

\vspace{0.2cm}
 
\noindent\ignorespaces \textbf{Keywords:} Poincar\'{e} chaos; Unpredictable function; Unpredictable solutions; Unpredictable sequence.

\vspace{0.6cm}

\section{Introduction}

It is useless to say that the theory of dynamical systems is a research of oscillations, and the latest motion of the theory is the Poisson stable trajectory \cite{p}. In paper \cite{Akhmet16} inspired by chaos investigation we have introduced a new type of oscillation, next to the Poisson stable one, and called the initial point for it the \textit{unpredictable point} and the trajectory  itself the \textit{unpredictable orbit}. These novelties make a connection of the homoclinic chaos and the latest types of chaos possible through the concept of \textit{unpredictability}, that is sensitivity assigned to a single orbit. Thus, the \textit{Poincar\'{e} chaos} concept has been eventually shaped in our paper \cite{Akhmet16}. We have also determined the unpredictable  function on the real axis as an unpredictable point of the Bebutov dynamics in paper \cite{Akhmet17} to involve widely differential and discrete equations to the chaos investigation. Nonetheless, we need a more precise description of what one understands as unpredictable function. The present research is devoted to this constructive duty. One can say that the analysis became productive, since we have learnt that the unpredictable functions can be bounded, and this is also true for Poisson functions, newly introduced in this paper. Thus, the discussion is focused on chaotic attractors bounded in the space variables. This is important for applications, and it remains legal through our suggestions.    

There  are different types of chaos. The first one is the homoclinic chaos \cite{sh}, which is generated from the famous manuscript of Poincar\'{e} \cite{p}. Another one is the Devaney chaos \cite{Devaney87} whose  ingredients are transitivity, Lorenz sensitivity \cite{Lorenz61} and the existence of infinitely many unstable periodic motions, which are dense in the chaotic attractor. The third one is the Li-Yorke chaos \cite{ly}, which is characterized by the presence of a scrambled set in which any pair of distinct points are proximal and frequently separated. One can see that the crucial difference of the homoclinic chaos from the others is the sensitivity and frequent separation features. When one says about homoclinic chaos it is assumed that there is the homoclinic structure, and moreover, there is instability. Lorenz \cite{Lorenz61} was the first who precised the  divergence of neighbor motions as sensitivity, a specific sort of instability and this was followed by Li and Yorke \cite{ly}, when they use frequent separation and proximality for the same purpose.
The absence of the quantitative description of instability makes homoclinic chaos different from  the late definitions, and this causes some sort of inconvenience. This is because the sensitivity is assumed to be one of the main ingredients of chaos in its modern comprehension. H. Poincar\'{e} himself was aware about the divergence of initially nearby trajectories, but had not given exact prescriptions how it should be proceeded. This is  why, one can say that the puzzle construction, which was initiated and designed by the French genius is not still completed. In our study, we are trying to make a contribution to this puzzle working. 
For that purpose we have utilized open Poisson stable motions, which accompany homoclinic chaos \cite{sh}. 
In paper \cite{Akhmet16} we developed the concept of Poisson stable point to the concept of unpredictable point utilizing \textit{unpredictability} as individual sensitivity for a motion. Thus, by issuing from the single point of a trajectory we use it as the Ariadne's thread to come to phenomenon, which we call as \textit{Poincar\'{e} chaos} in \cite{Akhmet16}. This phenomenon makes the all types of chaos closer, since it is another description of motions in dynamics with homoclinic structure and from another side it admits ingredients similar to late chaos types. That is, transitivity, sensitivity, frequent separation and proximality. Presence of infinitely many periodic motions in late definitions can be substituted by continuum of Poisson stable orbits. Our main hopes are that this suggestions may bring research of chaos back to the theory of classical dynamical systems.
The strong  argument for this, is the fact that we introduced a new type  of motions. That is the already existing list of oscillations in dynamical systems from equilibrium to Poisson stable orbits is now prolonged with unpredictable motions. This enlargement will give a push for the further extension of dynamical systems theory. In applications, some properties and/or laws of dynamical systems can be lost or ignored. For example, if one considers non-autonomous or non-smooth systems. Then we can apply unpredictable functions \cite{Akhmet17}, a new type of oscillations which immediately follow almost periodic solutions of differential equations in the row of bounded solutions. They can be investigated for any type of equations, since by our results they can be treated by methods of qualitative theory of differential equations.

We utilize the topology of uniform convergence on any compact subset of the real axis to introduce the unpredictable functions. More precisely, the Bebutov dynamical system \cite{sell} has been applied. Additionally, using the same dynamics we have introduced Poisson functions. All these make our duty of incorporating chaos investigation to theory of differential equations initiated in papers \cite{Akh3}-\cite{Akh15} and in the book \cite{Akh14} seems to proceed in the correct way.

The main goal of this paper is the construction of a concrete unpredictable function and its application to differential equations. To give the procedure we start with unpredictable sequences as motions of symbolic dynamics and the logistic map. Then an unpredictable function is determined as an improper convolution integral with a relay function. Finally, we demonstrated in examples  unpredictable functions as solutions of differential equations.

\section{Preliminaries}

Let $(X,d)$ be a metric space and $\pi: \mathbb T_+ \times X \to X,$ where $\mathbb T_+$ is either the set of non-negative real numbers or the set of non-negative integers, be a semi-flow on $X,$ i.e., $\pi(0,x)=x$ for all $x\in X,$ $\pi(t,x)$ is continuous in the pair of variables $t$ and $x,$ and $\pi(t_1,\pi(t_2,x))=\pi(t_1+t_2,x)$ for all $t_1, \ t_2 \in \mathbb T_+,$ $x\in X.$

A point $x \in X$ is called positively Poisson stable (stable $P^+$) if there exists a sequence $\left\{t_n\right\}$ satisfying $t_n \to \infty$ as $n \to \infty$ such that $\displaystyle \lim_{n\to \infty} \pi(t_n,x) = x$ \cite{Nemytskii60}. For a given point $x \in X,$ let us denote by $\Theta_x$ the closure of the trajectory $T(x)=\left\{\pi(t,x)  : \ t \in  \mathbb T_+ \right\},$ i.e., $\Theta_x = \overline{T(x)}.$ The set $\Theta_x$ is a quasi-minimal set if the point $x$ is stable $P^+$ and  $T(x)$ is contained in a compact subset of $X$ \cite{Nemytskii60}.

It was demonstrated by Hilmy \cite{Hilmy36} that if the trajectory corresponding to a Poisson stable point $x$ is contained in a compact subset of $X$ and it is neither a rest point nor a cycle, then the quasi-minimal set contains an uncountable set of motions everywhere dense and Poisson stable.  The following theorem can be proved by adapting the technique given in \cite{Nemytskii60,Hilmy36}.

\begin{theorem} (\cite{Akhmet16}) \label{Hilmy_thm}
Suppose that $x\in X$ is stable $P^+$ and $T(x)$ is contained in a compact subset of $X.$ If $\Theta_x$ is neither a rest point nor a cycle, then it contains an uncountable set of motions everywhere dense and stable $P^+.$
\end{theorem}

The definitions of an unpredictable point and unpredictable trajectory are as follows.

\begin{definition} (\cite{Akhmet16}) \label{denf_unpredictable_traj}
A point $x \in X$ and the trajectory through it are unpredictable if there exist a positive number $\epsilon_0$ (the unpredictability constant) and sequences $\left\{t_n\right\}$ and $\left\{\tau_n\right\},$ both of which diverge to infinity, such that $\displaystyle \lim_{n\to\infty} \pi(t_n,x)=x$ and $d(\pi(t_n+\tau_n,x),\pi(\tau_n,x)) \ge \epsilon_0$ for each $n \in \mathbb N.$
\end{definition}

Markov \cite{Nemytskii60} proved that a trajectory stable in both Poisson and Lyapunov (uniformly) senses must be an almost periodic one. Since Definition \ref{denf_unpredictable_traj} implies instability, an unpredictable motion cannot be almost periodic. In particular, it is neither an equilibrium, nor a cycle.

Based on unpredictable points, a new chaos definition was provided in the paper \cite{Akhmet16} as follows.

\begin{definition} (\cite{Akhmet16})
The dynamics on the quasi-minimal set $\Theta_x$ is called Poincar\'{e} chaotic if $x$ is an unpredictable point.
\end{definition}

It is worth noting that Poincar\'{e} chaos admits properties similar to the ingredients of Devaney chaos \cite{Devaney87}. In the paper \cite{Akhmet16}, it was proved that if $x$ is an unpredictable point, then the dynamics on $\Theta_x$ is sensitive. That is, there exists a positive number $\widetilde{\epsilon}_0$ such that for each $x_1\in \Theta_x$ and for each positive number $\delta$ there exist a point $x_2\in \Theta_x$ and a positive number $\bar{t}$ such that $d(x_1,x_2)<\delta$ and $d(f(\bar{t},x_1),f(\bar{t},x_2)) \ge \widetilde{\epsilon}_0.$ Besides, since the trajectory $T(x)$ is dense in $\Theta_x,$ transitivity is also present in the dynamics. Moreover, according to Theorem \ref{Hilmy_thm}, there exists a continuum of stable $P^+$ orbits in $\Theta_x.$ In our previous paper \cite{Akhmet16} we gave the definition of Poincar\'{e} chaos for flows, but in this paper we give the definition for semi-flows, since the discussion in the paper \cite{Akhmet16} is valid also for the latter case.

Let us denote by $C(\mathbb{R})$ the set of continuous functions defined on $\mathbb{R}$ with values in $\mathbb{R}^m,$ and assume that $C(\mathbb{R})$ has the topology of uniform convergence on compact sets, i.e., a sequence $\{h_k \}$ in $C(\mathbb{R})$ is said to converge to a limit $h$ if for every compact set $\mathcal{U}\subset \mathbb{R}$ the sequence of restrictions $\{h_k|_{\mathcal{U}} \}$ converges to $\{h|_{\mathcal{U}} \}$ uniformly.

One can define a metric $\rho$ on $C(\mathbb{R})$ as \cite{sell}
\begin{eqnarray} \label{distance_formula}
\rho(h_1,h_2)=\sum_{k=1}^{\infty}{2^{-k}\rho_k(h_1,h_2)},
\end{eqnarray}
where $h_1,$ $h_2$ belong to $C(\mathbb{R})$ and $$\rho_k(h_1,h_2)=\min \bigg\{ 1,\sup_{s\in [-k,k]}\| h_1(s)-h_2(s)\| \bigg\}, \ k\in \mathbb N.$$

Let us define the mapping $\pi:\mathbb{R}_+ \times C(\mathbb{R})\to C(\mathbb{R})$ by $\pi(t,h)=h_t,$ where $h_t(s)=h(t+s).$ The mapping $\pi$ is a semi-flow on $C(\mathbb{R}),$ and it is called the Bebutov dynamical system \cite{sell}.
 	
Using the Bebutov dynamical system, we give the descriptions of a Poisson function and an unpredictable function in the next definitions. 
 	
\begin{definition}   \label{poisson_defn}
A Poisson function is a Poisson stable point of the Bebutov dynamical system.
\end{definition}

\begin{definition} (\cite{{Akhmet17}}) \label{unpredict_defn}
An unpredictable function	is an unpredictable point of the Bebutov dynamical system.
\end{definition}	

It is clear that an unpredictable function is a Poisson function. According to Theorem $III.3$ \cite{sell}, a motion $\pi(t,h)$ lies in a compact set if $h$ is a bounded and uniformly continuous function. Therefore, an unpredictable function $h$ determines Poincar\'{e} chaos in the Bebutov dynamical system if it is bounded and uniformly continuous. Moreover, any system of differential equations which admits uniformly continuous and bounded unpredictable solution has a Poincar\'{e} chaos. For differential equations, we say that a solution is an unpredictable one if it is uniformly continuous and bounded on the real axis.

Let us consider the system 
\begin{eqnarray} \label{sys_unpredict_exten}
x'(t)=Ax(t)+f(x(t))+g(t),
\end{eqnarray}
where all eigenvalues of the constant $p\times p$ matrix $A$ have negative real parts, the function $f:\mathbb R^p \to\mathbb R^p$ is bounded, and the function $g:\mathbb R \to \mathbb R^p$ is a uniformly continous and bounded.  Since the eigenvalues of the matrix $A$ have negative real parts, there exist positive numbers $K_0$ and $\omega$ such that $\left\|e^{At}\right\|\le K_0 e^{\omega t}$ for $t\ge 0$ \cite{Hale80}.

The presence of an unpredictable solution in the dynamics of (\ref{sys_unpredict_exten}) is mentioned in the next theorem.

\begin{theorem} (\cite{{Akhmet17}}) \label{thm_unpredict_exten}
If $g(t)$ is an unpredictable function and the function $f(x)$ is Lipschitzian with a sufficiently small Lipschitz constant $L_f$ such that $K_0 L_f - \omega <0$, then system (\ref{sys_unpredict_exten}) possesses a unique uniformly exponentially stable unpredictable solution.
\end{theorem}
 
In the next two sections we will consider unpredictable functions whose domain consists of   all integers,  that  is, \textit{unpredictable sequences}.

\section{An unpredictable sequence  of the symbolic dynamics}
 
In this section, we will show the presence of an unpredictable point in the symbolic dynamics \cite{Devaney87,Wiggins88} with a distinguishing feature.

Let us consider the space $\displaystyle \Sigma_2=\left\{s=(s_0s_1s_2\ldots) \ | \ s_j =0 \ \textrm{or} \ 1 \right\}$ of infinite sequences of $0$'s and $1$'s with the metric $$d(s,t)=\displaystyle \sum_{k=0}^{\infty} \frac{\left|s_k-t_k\right|}{2^k},$$ where $s=(s_0s_1s_2\ldots),$ $t=(t_0t_1t_2\ldots)\in \Sigma_2.$ The Bernoulli shift $\sigma: \Sigma_2 \to \Sigma_2$ is defined as $\sigma(s_0s_1s_2\ldots)=(s_1s_2s_3\ldots).$ The map $\sigma$ is continuous and $\Sigma_2$ is a compact metric space \cite{Devaney87,Wiggins88}.

In the book \cite{Devaney87}, the sequence 
\begin{eqnarray} \label{sequence_s_star}
s^* = (\underbrace{0 \ 1}_{1 \, blocks}|\underbrace{00 \ 01 \ 10 \ 11}_{2 \,  blocks}|\underbrace{000 \ 001 \ 010 \ 011 \ \ldots}_{3 \, blocks}| \ldots),
\end{eqnarray}
was considered, which is constructed by successively listing all blocks of $0$'s and $1$'s of length $n,$ then length $n+1,$ etc. In the proof of the next lemma, an element $s^{**}=(s^{**}_0 s^{**}_1 s^{**}_2 \ldots)$ of $\Sigma_2$ will be constructed in a similar way to $s^*$ with the only difference that the order of the blocks will be chosen in a special way. An extension of the constructed sequence to the left hand side will also be provided.

\begin{lemma} \label{symbolic_dyn_lemma}
For each increasing sequence $\left\{m_n\right\}$ of positive integers, there exist a sequence $s^{**} \in \Sigma_2$ and sequences $\left\{\alpha_n\right\},$ $\left\{\beta_n\right\}$ of positive integers, both of which diverge to infinity, such that 
\begin{enumerate}
\item[\textbf{(i)}] $d(\sigma^{\alpha_n+r}(s^{**}),\sigma^r(s^{**})) \le 2^{-m_n}, \ r=-n,-n+1,\ldots,n,$
\item[\textbf{(ii)}] $d(\sigma^{\alpha_n+\beta_n}(s^{**}),\sigma^{\beta_n}(s^{**})) \ge 1$ for each $n\in\mathbb N.$
\end{enumerate}
\end{lemma}

\noindent \textbf{Proof.} Fix an arbitrary increasing sequence $\left\{m_n\right\}$ of positive integers. For each $n\in\mathbb N,$ define $\alpha_n=\sum_{k=1}^{n+m_n} k2^k$ and $\beta_n=n+m_n+1.$ Clearly, both of the sequences $\left\{\alpha_n\right\}$ and $\left\{\beta_n\right\}$ diverge to infinity. We will construct a sequence $s^{**} \in \Sigma_2$ such that the inequalities $(i)$ and $(ii)$ are valid.


  
First of all, we choose the terms $s^{**}_0, s^{**}_1, \ldots, s^{**}_{\alpha_1-1}$ by successively placing the blocks of $0$'s and $1$'s in an increasing length, starting from the blocks of length $1$ till the end of the ones with length $m_1+1.$ The order of the blocks with the same length can be arbitrary without any repetitions. Let us take $s^{**}_{\alpha_1+k}=s^{**}_k$ for $k=0,1,\ldots,m_1+1,$ i.e., the terms of the first block $(s^{**}_{\alpha_1} s^{**}_{\alpha_1+1} \cdots s^{**}_{\alpha_1+m_1+1})$ of length $m_1+2$ is chosen the same as the first $m_1+2$ terms of the sequence $s^{**}.$ Moreover, we take the second block of length $m_1+2$ in a such a way that its first term $s^{**}_{\alpha_1+\beta_1}$ is different from $s^{**}_{\beta_1}.$ After that we continue placing the remaining blocks of length $m_1+2$ and the ones with length greater than $m_1+2$ till the last block of length $m_2+2,$ and again, the blocks of the same length can be in any order without repetitions.

For each $n \ge 2,$ we set the last block of length $m_n+n$ such that $s^{**}_{\alpha_n-k}=s^{**}_{\alpha_k-k}$ for each $k=1,2,\ldots,n-1.$ Then, the terms of the first block of length $m_n+n+1$ are constituted by taking $s^{**}_{\alpha_n+k}=s^{**}_k,$ $k=0,1,\ldots,m_n+n.$ Moreover, the second block of length $m_n+n+1$ is chosen such that $s^{**}_{\alpha_n+\beta_n} \neq s^{**}_{\beta_n}.$ Lastly, the remaining blocks of $0$'s and $1$'s are successively placed similar to the case mentioned above so that the lengths of the blocks in $s^{**}$ are in an increasing order and there are no repetitions within the blocks of the same length. By this way the construction of the sequence $s^{**}\in\Sigma_2$ is completed. We fix the extension of the sequence $s^{**}$ to the left by choosing $\sigma^{-k}(s^{**})_0=s^{**}_{\alpha_k-k},$ $k\in\mathbb N.$ For each $n\in\mathbb N,$ we have $\sigma^{\alpha_n-n}(s^{**})_k = \sigma^{-n}(s^{**})_k,$ $k=0,1,\ldots,m_n+2n$ and $\sigma^{\alpha_n+\beta_n}(s^{**})_0\neq \sigma^{\beta_{n}}(s^{**})_0$ so that the inequalities $(i)$ and $(ii)$ are valid. $\square$

The technique presented in \cite{Devaney87} can be used to show that the trajectory $T(s^{**})=\left\{\sigma^i(s^{**}): \ i\in\mathbb Z\right\}$ is dense in $\Sigma_2,$ i.e., $\Theta_{s^{**}}=\Sigma_2.$ By Lemma 2.2 in \cite{Akhmet16} any sequence $\sigma^i(s^{**}),$ $i\in \mathbb Z,$ is an unpredictable point of the Bernoulli dynamics on $\Sigma_2,$ and $\Sigma_2$ is a quasi-minimal set. It implies from the last theorem that $T(s^{**})$ is an unpredictable function on $\mathbb Z,$ i.e., an unpredictable sequence. According to Theorem 3.1 presented in \cite{Akhmet16}, the dynamics on $\Sigma_2$ is Poincar\'{e} chaotic. Moreover, there are infinitely many unpredictable sequences in the set.

\section{An unpredictable solution of the logistic map} \label{Poincaresec4}

In this section, we will demonstrate the presence of an unpredictable solution of the equation 
\begin{eqnarray} \label{logisticc_map}
\eta_{n+1}=F_{\mu}(\eta_n),
\end{eqnarray} 
where $F_{\mu}(s)=\mu s (1-s)$ is the logistic map.

The  result is provided in the next theorem.

\begin{theorem} \label{logisticc_theorem} 
For each $\mu \in [3+(2/3)^{1/2}, 4]$ and sequence of positive numbers $\delta_n \to 0,$ there exists a solution $\left\{\eta_n\right\},$ $n \in \mathbb Z,$ of equation (\ref{logisticc_map}) such that 
\begin{enumerate}
\item[\textbf{(i)}] $\left|\eta_{i_n+r} - \eta_r\right| < \delta_n,$ $r=-h_0 n, -h_0 n+1, \ldots, h_0 n,$ 
\item[\textbf{(ii)}] $\left|\eta_{i_n+j_n}-\eta_{j_n}\right|\ge\epsilon_0$ for each $n \in \mathbb N,$
\end{enumerate}
where $\epsilon_0$ is a positive number, $h_0 >4$ is a natural number, and $\left\{i_n\right\},$ $\left\{j_n\right\}$ are integer valued sequences both of which diverge to infinity.
\end{theorem}
 
\noindent \textbf{Proof.} 
Fix $\mu \in [3+(2/3)^{1/2}, 4]$ and a sequence $\left\{\delta_n\right\}$ of positive real numbers with $\delta_n \to 0$ as $n\to \infty.$ Take a neighborhood $U\subset [0,1]$ of the point $1-1/\mu.$ According to Theorem $6$ of paper \cite{Shi07}, there exist a natural number $h_0 > 4$ and a Cantor set $\Lambda \subset U$ such that the map $F_{\mu}^{h_0}$ on $\Lambda$ is topologically conjugate to the Bernoulli shift $\sigma$ on $\Sigma_2.$ Therefore, there exists a homeomorphism $S: \Sigma_2 \to \Lambda$ such that $S \circ \sigma = F^{h_0}_{\mu} \circ S.$ Since $S$ is uniformly continuous on $\Sigma_2,$ for each $n\in\mathbb N,$ there exists a number $\overline{\delta}_n>0$ such that for each $s^1,$ $s^2\in \Sigma_2$ with $d(s^1,s^2)<\overline{\delta}_n,$ we have $\left|\eta^1-\eta^2\right|<\delta_n / \mu^{h_0-1},$ where $\eta^1=S(s^1),$ $\eta^2=S(s^2).$


Let $\left\{m_n\right\}$ be an increasing sequence of natural numbers such that $2^{-m_n}<\overline{\delta}_n$ for each $n\in\mathbb N.$
According to Lemma \ref{symbolic_dyn_lemma}, there exist a sequence $s^{**} \in \Sigma_2$ and sequences $\left\{\alpha_n\right\},$ $\left\{\beta_n\right\}$ both of which diverge to infinity such that 
$d(\sigma^{\alpha_n+r}(s^{**}),\sigma^r(s^{**})) \le 2^{-m_n},$ $r=-n,-n+1,\ldots,n,$ and $d(\sigma^{\alpha_n+\beta_n}(s^{**}),\sigma^{\beta_n}(s^{**})) \ge 1$ for each $n\in\mathbb N.$

Now, let $\left\{\eta_n\right\},$ $n\in \mathbb Z,$ be the solution of (\ref{logisticc_map}) with $\eta_{h_0 k}=S(\sigma^k(s^{**})),$ $k\in \mathbb Z.$ Since the inequality $\left|F_{\mu}(u_1)-F_{\mu}(u_2)\right| \le \mu \left|u_1-u_2\right|$ is valid for every $u_1,u_2\in [0,1],$ we have for each $n\in\mathbb N$ that $\left|\eta_{i_n+r} - \eta_r\right| < \delta_n,$ $r=-h_0 n, -h_0 n+1, \ldots, h_0 n,$ where $i_n=h_0 \alpha_n.$ Besides, using the arguments presented in \cite{Banks}, one can verify the existence of a positive number $\epsilon_0$ such that $\left|\eta_{i_n+j_n}-\eta_{j_n}\right|\ge\epsilon_0,$ $n \in \mathbb N,$ where $j_n=h_0 \beta_n.$ $\square$

By the topological equivalence and results on $\Sigma_2$ of the last section, one can make several observations from the proved theorem. Any number $\eta_n,$ $n\in \mathbb Z,$ is an unpredictable point of the logistic map dynamics, and the Cantor set $\Lambda$ mentioned in the proof of Theorem \ref{logisticc_theorem} is a quasi-minimal set. Moreover, the sequence $\left\{\eta_n\right\}$ is unpredictable. By Theorem 3.1 mentioned in \cite{Akhmet16}, the dynamics on the quasi-minimal set is Poincar\'{e} chaotic, and there are infinitely many unpredictable sequences in the set. The last observation will be applied in the next section to construct an  unpredictable function. 

\section{An unpredictable function} \label{Poincaresec5}

In this part of the paper, we will provide an example of an unpredictable function benefiting from the   dynamics  of the  logistic map (\ref{logisticc_map}).

Let us fix two different points $d_1$ and $d_2$ in $\mathbb R^p,$ and suppose that $\gamma$ is a positive number. Take a sequence $\left\{k_n\right\}$ of positive integers satisfying both of the inequalities $\displaystyle 2^{-k_n} \le \frac{1}{2n}$ and $\displaystyle e^{-\gamma k_n} \le \frac{\gamma}{4\left\|d_1-d_2\right\|n}$ for each $n\in \mathbb N.$ Fix $\mu \in [3+(2/3)^{1/2}, 4]$ and a sequence $\left\{\delta_n\right\}$ of positive numbers such that $\delta_n \le \displaystyle \frac{1}{12\left\|d_1-d_2\right\| n k_n},$ $n\in\mathbb N.$ In a similar way to the items $(i)$ and $(ii)$ of Theorem \ref{logisticc_theorem}, one can verify that there exist a positive number $\epsilon_0,$ a sequence $\left\{i_n\right\}$ of even positive integers, a sequence $\left\{j_n\right\}$ of positive integers, and a solution $\left\{\eta_n\right\},$ $n\in\mathbb Z,$ of the logistic map (\ref{logisticc_map}) such that the inequalities
\begin{eqnarray}  \label{ineqq1}
\left|\eta_{i_n+r} - \eta_r\right| \le \delta_n, \ r=-2k_n, -2k_n+1, \ldots, k_n-1,
\end{eqnarray}
and 
\begin{eqnarray}  \label{ineqq2}
\left|\eta_{i_n+j_n} - \eta_{j_n}\right| \ge \epsilon_0
\end{eqnarray} 
hold for each $n\in\mathbb N.$ 

It is easy to observe that the constructed sequence $\left\{\eta_{n}\right\}$ is unpredictable and consequently, it generates a quasi-minimal set and Poincar\'{e} chaos similar to that of the last section.

Now, consider the function $\phi: \mathbb R \to \mathbb R^p$ defined as 
\begin{eqnarray} \label{func_phi}
\phi(t) = \displaystyle \int\displaylimits^{t}_{-\infty} e^{-\gamma (t-s)} \nu(s) ds,
\end{eqnarray}
where the function $\nu(t)$ is defined as
\begin{eqnarray*} \label{func_relay}
\nu(t) = \left\{\begin{array}{ll} d_1, ~\textrm{if} & \zeta_{2j} < t  \leq \zeta_{2j+1}, \ j \in\mathbb Z, \\
                                d_2, ~\textrm{if} & \zeta_{2j-1} < t \leq \zeta_{2j}, \ j \in\mathbb Z,
\end{array} \right.
\end{eqnarray*}
and the sequence $\left\{\zeta_j\right\},$ $j\in\mathbb Z,$ of switching moments is defined through the equation $\zeta_j = j + \eta_j$ for each $j,$ in which $\left\{\eta_j\right\}$ is the solution of (\ref{logisticc_map}) satisfying (\ref{ineqq1}) and (\ref{ineqq2}). The function $\phi(t)$ is bounded such that $\displaystyle \sup_{t\in\mathbb R} \left\|\phi(t)\right\|\le \frac{\max\left\{\left\|d_1\right\|,\left\|d_2\right\|\right\}}{\gamma}.$ Moreover, $\phi(t)$ is uniformly continuous since its derivative is bounded.

In the proof of the following theorem, we will denote by $\widehat{(a,b]}$ the oriented interval such that $\widehat{(a,b]}=(a,b]$ if $a < b$ and $\widehat{(a,b]}=(b,a]$ if $a > b.$

\begin{theorem} \label{thm_unpredictable}
The function $\phi(t)$ is unpredictable.
\end{theorem} 
 
\noindent \textbf{Proof.} First of all, we will show that $\rho(\phi_{i_n},\phi) \to 0$ as $n\to\infty,$ where $\rho$ is the metric defined by equation (\ref{distance_formula}). Let us fix an arbitrary natural number $n.$ The functions $\phi(i_n+s)$ and $\phi(s)$ satisfy the equation
\begin{eqnarray*}
\phi(i_n+s)-\phi(s)=\displaystyle \int\displaylimits^{s}_{-\infty} e^{-\gamma (s-u)} \left( \nu(i_n+u) - \nu(u) \right) du.
\end{eqnarray*}
It is worth noting that for each $r\in\mathbb Z$ both of the points $\zeta_r$ and $\zeta_{i_n+r}-i_n$ belong to the interval $(r,r+1).$ Moreover, $\left\|\nu(i_n+s)-\nu(s)\right\|=\left\|d_1-d_2\right\|$ for $s\in \displaystyle \bigcup_{r=-\infty}^{\infty} \widehat{(\zeta_r,\zeta_{i_n+r}-i_n]},$ and $\left\|\nu(i_n+s)-\nu(s)\right\|=0,$ otherwise.

Since for each $r=-2k_n,-2k_n+1,\ldots,k_n-1$ the distance between the points $\zeta_r$ and $\zeta_{i_n+r}-i_n$ are at most $\delta_n,$
one can verify for each $s \in [-k_n,k_n]$ that
\begin{eqnarray*}
&\left\|\phi(i_n+s)-\phi(s)\right\| & \le \displaystyle \int\displaylimits^{-2k_n}_{-\infty} e^{-\gamma (s-u)} \left\| \nu(i_n+u) - \nu(u) \right\| du 
 + \displaystyle \sum^{k_n-1}_{r=-2k_n} \bigg| \int\displaylimits_{\zeta_{i_n+r}-i_n}^{\zeta_r} \left\| \nu(i_n+u) - \nu(u) \right\| du \bigg| \\
&& \le \displaystyle \frac{\left\|d_1-d_2\right\|}{\gamma} e^{-\gamma k_n} + 3 k_n \delta_n \left\|d_1-d_2\right\| \\
&& \le \displaystyle \frac{1}{2n}. 
\end{eqnarray*}
Hence, we have
\begin{eqnarray*}
\rho(\phi_{i_n},\phi)  = \displaystyle \sum_{k=1}^{\infty} 2^{-k} \rho_k(\phi_{i_n},\phi) 
 \le \displaystyle \sum_{k=1}^{k_n} 2^{-k} \rho_k(\phi_{i_n},\phi) + 2^{-k_n} 
 \le \displaystyle \frac{1}{n}.
\end{eqnarray*}
The last inequality implies that $\rho(\phi_{i_n},\phi) \to 0$ as $n \to \infty,$ i.e., $\phi(t)$ is a Poisson function.

Now, let us show the existence of a positive number $\overline{\epsilon}_0$ satisfying $\overline{\epsilon}_0 \to 0$ as $\epsilon_0 \to 0$  such that $\rho(\phi_{i_n+j_n},\phi_{j_n}) \ge \overline{\epsilon}_0$ for each $n \in \mathbb N.$ For a fixed natural number $n,$ using the equations
\begin{eqnarray*}
\phi(i_n+j_n+s)=e^{-\gamma s} \phi(i_n+j_n) + \displaystyle \int\displaylimits^{s}_{0} e^{-\gamma (s-u)} \nu(i_n+j_n+u) du
\end{eqnarray*}
and
\begin{eqnarray*}
\phi(j_n+s)=e^{-\gamma s} \phi(j_n) + \displaystyle \int\displaylimits^{s}_{0} e^{-\gamma (s-u)} \nu(j_n+u) du,
\end{eqnarray*}
we obtain that
\begin{eqnarray*}
&\left\|\phi(i_n+j_n+1) - \phi(j_n+1) \right\| & \ge  \displaystyle \bigg\| \int\displaylimits_{\zeta_{i_n+j_n}-i_n-j_n}^{\zeta_{j_n}-j_n} e^{-\gamma (1-u)} (d_1-d_2) du \bigg\| - e^{-\gamma} \left\|\phi(i_n+j_n)-\phi(j_n) \right\| \\
&& \ge  \displaystyle \frac{e^{\gamma \epsilon_0} - 1}{\gamma e^{\gamma}} \left\|d_1-d_2\right\|- e^{-\gamma} \left\|\phi(i_n+j_n)-\phi(j_n) \right\|.
\end{eqnarray*}
Therefore, it can be verified for each $k\in\mathbb N$ that
\begin{eqnarray} \label{unp_proof_ineqq1}
 \displaystyle  \sup_{s \in [-k,k]} \left\|\phi(i_n+j_n+s) - \phi(j_n+s) \right\| \ge \frac{\left(e^{\gamma \epsilon_0}-1\right) \left\|d_1-d_2\right\|}{\gamma \left(1+e^{\gamma}\right)}.
\end{eqnarray}
Let us denote $$\overline{\epsilon}_0 = \min\left\{1,\frac{\left(e^{\gamma \epsilon_0}-1\right) \left\|d_1-d_2\right\|}{\gamma \left(1+e^{\gamma}\right)}\right\}.$$
It can be confirmed by means of inequality (\ref{unp_proof_ineqq1}) that $\rho_k(\phi_{i_n+j_n},\phi_{j_n}) \ge \overline{\epsilon}_0,$ $k\in\mathbb N.$ Thus, $\rho(\phi_{i_n+j_n},\phi_{j_n}) \ge \overline{\epsilon}_0$ for each $n\in\mathbb N.$
Consequently, the function $\phi(t)$ is unpredictable. $\square$


One of the possible ways useful for applications to generate unpredictable functions from a given one is provided in the next theorem.

\begin{theorem} \label{lemma_unpredictablee}
Let $\phi: \mathbb R \to \mathscr{H}$ be an unpredictable function, where $\mathscr{H}$ is a bounded subset of $\mathbb R^p.$ If $h:\mathscr{H} \to \mathbb R^q$ is a function such that there exist positive numbers $L_1$ and $L_2$ satisfying $L_1 \left\|u-\overline{u}\right\| \le \left\|h(u)-h(\overline{u})\right\|\le L_2 \left\|u-\overline{u}\right\|$ for all $u, \ \overline{u} \in \mathscr{H},$ then the function $\psi:\mathbb R \to \mathbb R^q$ defined as $\psi(t)=h(\phi(t))$ is also unpredictable.
\end{theorem}

\noindent \textbf{Proof.} Since $\phi(t)$ is an unpredictable function, there exist a positive number $\epsilon_0$ and sequences $\left\{t_n\right\}$ and $\left\{\tau_n\right\},$ both of which diverge to infinity, such that $\displaystyle \lim_{n \to \infty} \rho(\phi_{t_n},\phi)=0$ and $\rho(\phi_{t_n+\tau_n},\phi_{\tau_n}) \ge \epsilon_0$ for each $n\in\mathbb N.$

Firstly, we will show that $\displaystyle \lim_{n \to \infty} \rho(\psi_{t_n},\psi)=0.$ Fix an arbitrary positive number $\epsilon,$ and let us denote $\alpha=\max \left\{1, L_2\right\}.$ There exists a natural number $n_0$ such that for all $n\ge n_0$ the inequality $\rho(\phi_{t_n},\phi) < \epsilon /\alpha$ is valid. For each $k \in \mathbb N,$ one can confirm that
\begin{eqnarray*}
&\displaystyle \rho_k(\psi_{t_n},\psi) & =  \min \bigg\{1, \sup_{s\in[-k,k]} \left\|h(\phi(t_n+s)) - h(\phi(s)) \right\| \bigg\} \\
&& \le \min \bigg\{1, L_2  \sup_{s\in[-k,k]} \left\|\phi(t_n+s) - \phi(s) \right\| \bigg\} \\
&& \le \alpha \rho_k (\phi_{t_n}, \phi).
\end{eqnarray*}
Therefore, it can be verified for each $n\ge n_0$ that 
\begin{eqnarray*}
\rho(\psi_{t_n},\psi) \le \alpha \rho(\phi_{t_n},\phi) < \epsilon.
\end{eqnarray*}
Hence, $\displaystyle \lim_{n \to \infty} \rho(\psi_{t_n},\psi)=0.$

Next, we will show the existence of a positive number $\overline{\epsilon}_0$ such that $\rho(\psi_{t_n+\tau_n},\psi_{\tau_n})\ge \overline{\epsilon}_0$ for each $n\in\mathbb N.$ Denote $\beta=\min\left\{1,L_1\right\}.$ For each $k\in\mathbb N,$ we have that 
\begin{eqnarray*}
&\rho_k(\psi_{t_n+\tau_n},\psi_{\tau_n}) & = \min \bigg\{1, \sup_{s\in[-k,k]} \left\|h(\phi(t_n+\tau_n+s)) - h(\phi(\tau_n+s)) \right\| \bigg\} \\
&& \displaystyle \ge  \min \bigg\{1, L_1 \sup_{s\in[-k,k]} \left\|\phi(t_n+\tau_n+s) - \phi(\tau_n+s) \right\| \bigg\} \\
&& \ge \beta \rho_k(\phi_{t_n+\tau_n},\phi_{\tau_n}).
\end{eqnarray*}
Thus, the inequality
\begin{eqnarray*}
\rho(\psi_{t_n+\tau_n},\psi_{\tau_n}) \ge \beta \rho(\phi_{t_n+\tau_n},\phi_{\tau_n}) \ge \overline{\epsilon}_0
\end{eqnarray*}
holds for each $n\in\mathbb N$, where $\overline{\epsilon}_0=\beta \epsilon_0.$ Consequently, the function $\psi(t)$ is unpredictable. $\square$

A corollary of Theorem \ref{lemma_unpredictablee} is as follows.

\begin{corollary} \label{corollary_unpredictablee}
If $\phi: \mathbb R \to \mathscr{H}$ is an unpredictable function, where $\mathscr{H}$ is a bounded subset of $\mathbb R^p,$ then the function $\psi:\mathbb R \to \mathbb R^p$ defined as $\psi(t)=P \phi(t),$ where $P$ is a constant, nonsingular, $p\times p$  matrix, is also an unpredictable function.
\end{corollary}

\noindent \textbf{Proof.} The function $h:\mathscr{H} \to \mathbb R^p$ defined as $h(u)=Pu$ satisfies the inequality $$L_1 \left\|u_1-u_2\right\| \le \left\|h(u_1)-h(u_2)\right\|\le L_2 \left\|u_1-u_2\right\|,$$ for $u_1,  u_2 \in \mathscr{H}$ with $L_1= 1/ \left\|P^{-1}\right\|$ and $L_2= \left\|P\right\|.$ Therefore, by Theorem \ref{lemma_unpredictablee}, the function $\psi(t)$ is unpredictable.  $\square$

In the next section, the existence of Poincar\'{e} chaos in the dynamics of differential equations will be presented.

\section{Unpredictable solutions of differential equations}

Consider the differential equation
\begin{eqnarray} \label{difeqnex1}
x'(t)=- \frac{3}{2} x(t) + \nu(t),
\end{eqnarray}
where the function $\nu(t)$ is defined as 
\begin{eqnarray} \label{example_relay1}
\nu(t) = \left\{\begin{array}{ll} ~0.7, ~~ \textrm{if} & \zeta_{2j} < t  \leq \zeta_{2j+1}, \ j \in\mathbb Z, \\
                                  -0.4, ~ \textrm{if} & \zeta_{2j-1} < t \leq \zeta_{2j}, \ j \in\mathbb Z.
\end{array} \right.
\end{eqnarray}
In (\ref{example_relay1}), the sequence $\left\{\zeta_j\right\}$ is defined as $\zeta_j=j+\eta_j,$ $j\in \mathbb Z,$   and  $\left\{\eta_j\right\}$ is   the  unpredictable sequence determined in  Section 
\ref{Poincaresec5}  for  the map (\ref{logisticc_map}) with $\mu=3.91.$

\begin{figure}[ht]
\centering
\includegraphics[width=14cm]{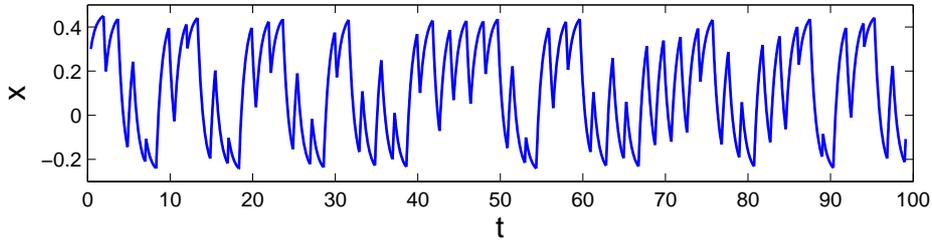}
\caption{Chaotic behavior in equation (\ref{difeqnex1}). The figure confirms that Poincar\'{e} chaos takes place in the dynamics of equation (\ref{difeqnex1}).}
\label{Poincare_fig1}
\end{figure}

According to Theorem \ref{thm_unpredictable}, 
\begin{eqnarray*} 
\psi(t) = \displaystyle \int\displaylimits^{t}_{-\infty} e^{-3 (t-s)/2} \nu(s) ds
\end{eqnarray*}
is a globally  asymptotically stable unpredictable solution  of (\ref{difeqnex1}).  We represent a solution of (\ref{difeqnex1}) corresponding to the initial data $x(\zeta_0)=0.3,$ $\zeta_0=0.4$  in Figure \ref{Poincare_fig1}. The choice  of the coefficient $\mu=3.91$ and the initial value $\zeta_0=0.4$ is approved by the shadowing analysis in paper \cite{Hammel87}. The simulation seen in Figure \ref{Poincare_fig1} supports the result of Theorem \ref{thm_unpredictable} such that the equation (\ref{difeqnex1}) behaves chaotically.

Next, we will demonstrate the chaotic behavior of a multidimensional system of differential equations.

Let us take into account the system
\begin{eqnarray} \label{example_systemm1}
x'(t)=Ax(t)+\nu(t),
\end{eqnarray} 
where 
$x(t)=(x_1(t), x_2(t), x_3(t))\in \mathbb R^3,$ $A=\left( \begin{array}{ccc}
-3 & 0 & -1 \\
-2 & 1 & -2 \\
-2 & 4 & -4 \end{array} \right),$
and the function $\nu:\mathbb R \to \mathbb R^3$ is defined as
\begin{eqnarray} \label{example_relay}
\nu(t) = \left\{\begin{array}{ll} (-1,1,2), ~\textrm{if} & \zeta_{2j} < t  \leq \zeta_{2j+1}, \ j \in\mathbb Z, \\
                                  (3,1,-1), ~\textrm{if} & \zeta_{2j-1} < t \leq \zeta_{2j}, \ j \in\mathbb Z.
\end{array} \right.
\end{eqnarray}
Similarly to equation (\ref{example_relay1}), in (\ref{example_relay}), the sequence $\left\{\zeta_j\right\}$ of switching moments is defined through the equation $\zeta_j=j+\eta_j,$ where $\left\{\eta_j\right\}$ is the unpredictable sequence determined in  Section \ref{Poincaresec5}  for  the map (\ref{logisticc_map}) with  $\mu=3.86.$

By means of the transformation $y=P^{-1}x,$ where 
$P=\left( \begin{array}{ccc}
1 & 2 & 1 \\
0 & 1 & -1 \\
-1 & 0 & -2 \end{array} \right),$
system (\ref{example_systemm1}) can be written as 
\begin{eqnarray} \label{systemm2}
y'(t)=Dy(t)+\overline{\nu}(t),
\end{eqnarray} 
where 
$D=\left( \begin{array}{ccc}
-2 & 0 & 0 \\
0 & -3 & 0 \\
0 & 0 & -1 \end{array} \right),$
and
\begin{eqnarray*} \label{example_relay2}
\overline{\nu}(t) = \left\{\begin{array}{ll} (0,0,-1), ~\textrm{if} & \zeta_{2j} < t  \leq \zeta_{2j+1}, \ j \in\mathbb Z, \\
                                  (1,1,0), ~~~\textrm{if} & \zeta_{2j-1} < t \leq \zeta_{2j}, \ j \in\mathbb Z.
\end{array} \right.
\end{eqnarray*}

System (\ref{systemm2}) admits an unpredictable solution $\overline{\psi}(t)$ in accordance with Theorem \ref{thm_unpredictable}. Therefore, Corollary \ref{corollary_unpredictablee} implies that  $\psi(t)=P \overline{\psi}(t)$ is an unpredictable solution of (\ref{example_systemm1}).
 
\begin{figure}[ht]
\centering
\includegraphics[width=14cm]{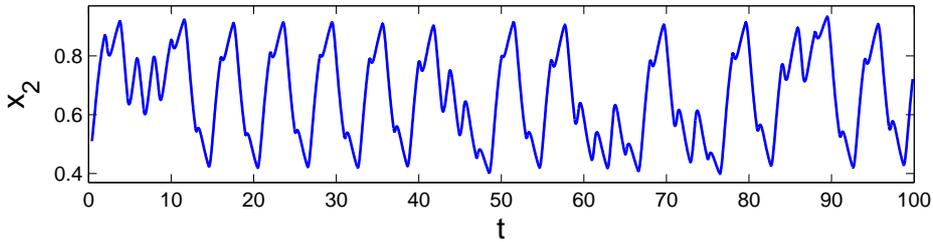}
\caption{The $x_2-$coordinate of the chaotic solution of system (\ref{example_systemm1}).}
\label{Poincare_fig2}
\end{figure}

\begin{figure}[ht]
\centering
\includegraphics[width=8cm]{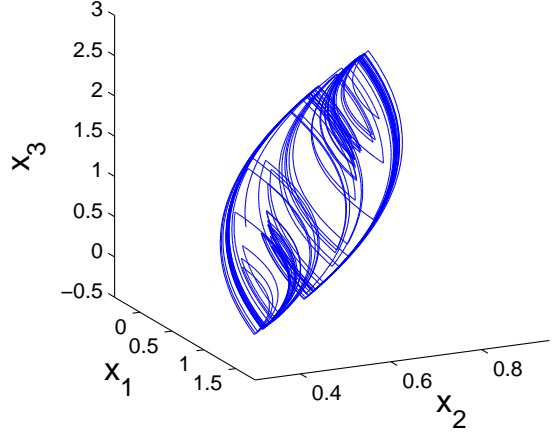}
\caption{Chaotic trajectory of system (\ref{example_systemm1}). The figure reveals the presence of Poincar\'{e} chaos in system (\ref{example_systemm1}).}
\label{Poincare_fig3}
\end{figure}

To demonstrate the chaotic behavior, we depict in Figure \ref{Poincare_fig2} the $x_2-$coordinate of the solution of (\ref{example_systemm1}) corresponding to the initial data $x_1(\zeta_0)=0.17,$ $x_2(\zeta_0)=0.51,$ $x_3(\zeta_0)=0.48,$ $\zeta_0=0.4.$ The coefficient $\mu=3.86$ and the initial value $\zeta_0=0.4$ are considered for shadowing in \cite{Hammel87}. The chaotic behavior is also valid in the remaining coordinates, which are not just pictured here. Moreover, Figure \ref{Poincare_fig3} shows the $3-$dimensional chaotic trajectory of the same solution. It is worth noting that the chaotic solutions of (\ref{example_systemm1}) take place inside the compact region
\begin{eqnarray} \label{Poincare_compact}
\mathscr{H}=\left\{(x_1,x_2,x_3) \in \mathbb R^3 \ | \ -1 \le x_1 \le 1.1, \ 0.3 \le x_2 \le 1, \ -0.4 \le x_3 \le 1.9 \right\}.
\end{eqnarray} 
Figures \ref{Poincare_fig2} and \ref{Poincare_fig3} support the results of Theorem \ref{thm_unpredictable} and Corollary \ref{corollary_unpredictablee} such that the represented solution behaves chaotically.
 
Now, we will demonstrate the extension of unpredictable solutions and Poincar\'{e} chaos. For that purpose, we consider the system
\begin{eqnarray} \label{extension_upredctible}
z'(t) = Bz(t) + f(z(t)) + h(\psi(t)),
\end{eqnarray}
where
$z(t)=(z_1(t), z_2(t), z_3(t))\in \mathbb R^3,$ 
$B=\left( \begin{array}{ccc}
-3 & 2 & -1 \\
0 & -5/2 & 0 \\
2 & -11/2 & -1 \end{array} \right),$
and $\psi(t)$ is the unpredictable solution of (\ref{example_systemm1}).
In system (\ref{extension_upredctible}), the function $f:\mathbb R^3 \to \mathbb R^3,$ $f(u)=(f_1(u),f_2(u),f_3(u)),$ is defined as $f_1(u)=0.03\sin u_1,$ $f_2(u)=0.04u_2^2$ for $\left|u_2\right| \le 1,$ $f_2(u)=0.04$ for $\left|u_2\right| > 1,$ $f_3(u)=0.06\tanh u_3,$ and the function $h:\mathbb R^3 \to \mathbb R^3,$ $h(u)=(h_1(u), h_2(u), h_3(u)),$ is defined as $h_1(u)=2\arctan(u_1),$ $h_2(u)=u_3+0.1 u_3^3,$ $h_3(u)=0.5 u_2^2,$ where $u=(u_1,u_2,u_3).$

The eigenvalues of the matrix $B$ are $-5/2,$ $-2 \pm i,$ and $e^{Bt}=Pe^{Jt}P^{-1},$ where
$J=\left( \begin{array}{ccc}
-5/2 & 0 & 0 \\
0 & -2 & -1 \\
0 & 1 & -2 \end{array} \right)$
and
$P=\left( \begin{array}{ccc}
1 & 0 & -1 \\
1/2 & 0 & 0 \\
1/2 & 1 & 1 \end{array} \right).$
One can verify that $\left\|e^{Bt}\right\| \le K_0 e^{-\omega t},$ $t\ge 0,$ with $K_0=\left\|P\right\| \left\|P^{-1}\right\|\approx 7.103$ and $\omega=2.$ The function $f(u)$ is bounded and it satisfies the Lipschitz condition $\left\|f(u)-f(\overline{u})\right\| \le L_f \left\|u-\overline{u}\right\|,$ $u, \ \overline{u} \in \mathbb R^3,$ with $L_f=0.08$ such that the inequality $K_0 L_f - \omega <0$ is valid.

On the other hand, the function $h(u)$ satisfies the conditions of Theorem \ref{lemma_unpredictablee} with $L_1=0.3$ and $L_2=2.083$ inside the region $\mathscr{H}$ defined by (\ref{Poincare_compact}) so that $h(x(t))$ is an unpredictable function if $x(t)$ is an unpredictable function. Therefore, according to Theorem \ref{thm_unpredict_exten}, system (\ref{extension_upredctible}) possesses a unique uniformly exponentially stable unpredictable solution.

\begin{figure}[ht]
\centering
\includegraphics[width=14cm]{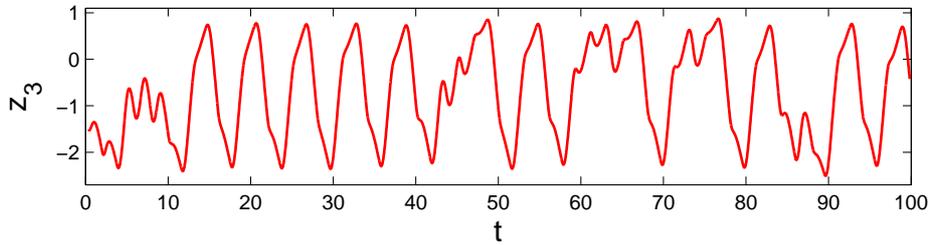}
\caption{The $z_3-$coordinate of the chaotic solution of system (\ref{Poincare_system3}). The figure manifests the presence of Poincar\'{e} chaos in system (\ref{Poincare_system3}).}
\label{Poincare_fig4}
\end{figure}

\begin{figure}[ht]
\centering
\includegraphics[width=8cm]{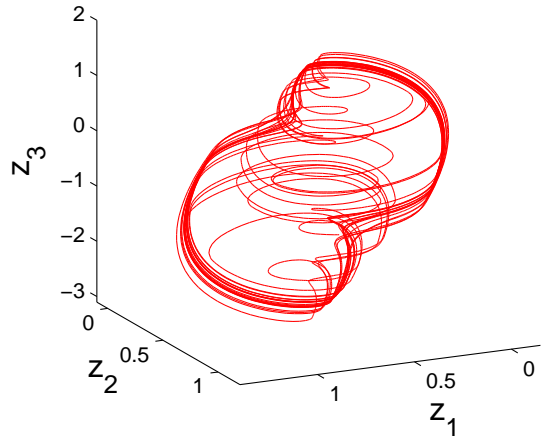}
\caption{Chaotic trajectory of system (\ref{Poincare_system3}). It is seen in the figure that the Poincar\'{e} chaos of system (\ref{example_systemm1}) is extended by (\ref{Poincare_system3}).}
\label{Poincare_fig5}
\end{figure}

Let us denote by $\theta(t)$ the solution of (\ref{example_systemm1}) whose trajectory is shown in Figure \ref{Poincare_fig3}.
To demonstrate the extension of Poincar\'{e} chaos, we take into account the system
\begin{eqnarray} \label{Poincare_system3}
z'(t)=Bz(t)+f(z(t))+h(\theta(t)).
\end{eqnarray}
We depict in Figure \ref{Poincare_fig4} the $z_3-$coordinate of the solution of (\ref{Poincare_system3}) with $z_1(\zeta_0)=0.53,$ $z_2(\zeta_0)=0.57,$ $z_3(\zeta_0)= -1.51,$ where $\zeta_0=0.4.$ Moreover, the trajectory of the same solution is represented in Figure \ref{Poincare_fig5}. The simulations shown in Figures \ref{Poincare_fig4} and \ref{Poincare_fig5} support Theorem \ref{thm_unpredict_exten} such that the Poincar\'{e} chaos of system (\ref{example_systemm1}) is extended by (\ref{Poincare_system3}).

\section{Conclusion}  

In our previous paper \cite{Akhmet16} we developed the concept of Poisson stable points to unpredictable ones, which imply the Poincar\'{e} chaos in the quasi-minimal set. This new definition makes homoclinic chaos and the late descriptions of the  phenomenon be closer to each other such that a unified background of chaos can be obtained in the future. Besides, in article \cite{Akhmet17}, an unpredictable function was defined as an unpredictable point of the Bebutov dynamical system. In the present paper, we have obtained samples of unpredictable functions and sequences, which are in the basis of Poincar\'{e} chaos. The unpredictable sequences are utilized in the construction of piecewise continuous functions, and such functions in their own turn are utilized for the construction of continuous unpredictable functions. Thus, one can claim that the basics of the new theory of unpredictable functions have been lied in the present research. Automatically, the results concerning the analyses of functions and sequences make it possible to formulate new problems of the existence of unpredictable solutions for differential equations of different  types as well as for discrete and hybrid systems of equations, similar to the results for periodic, almost periodic and other types of solutions. These are all strong arguments for the insertion of chaos research to the theory of differential equations. In addition to the role of the present paper for the theory of differential equations, the concept of unpredictable points and orbits introduced in our studies \cite{Akhmet16,Akhmet17} and the additional results of the present study will bring the chaos research to the scope of the classical theory of dynamical systems. Moreover, not less importantly, these concepts extend the boundaries of the theory of dynamical systems significantly, since we are dealing with a new type of motions, which are behind and next to Poisson stable trajectories. From another point of view, our study requests the development of techniques to determine whether a point is an unpredictable one in concrete dynamics. For that purpose one can apply the research results which exist for the indication of Poisson stable points \cite{sh}. One more interesting study depending on the present results can be performed if one tries to find a numerical approach for the recognition of unpredictable points.

\end{document}